\documentclass{article}
\usepackage[T1]{fontenc}
\usepackage[utf8]{inputenc}
\usepackage[absolute,overlay]{textpos}
\usepackage{hyperref}
\usepackage{authblk}
\usepackage[numbers,sort&compress,square]{natbib}
\usepackage{blindtext}
\usepackage[symbol]{footmisc}
\usepackage{graphicx}
\usepackage{amssymb}
\usepackage{titlesec}
\usepackage[top=1.5in, left=1.5in]{geometry}
\usepackage{endnotes}
\usepackage{xcolor}

\renewcommand{\emph}[1]{\textit{#1}}

\title{Model of cunning agents}

\author{Mateusz Denys\thanks{\noindent Email address: \href{mailto:mateusz.denys@gmail.com}{mateusz.denys@gmail.com}} }

\affil{AstroCeNT --- Particle Astrophysics Science and Technology Centre International Research Agenda, Nicolaus Copernicus Astronomical Center, Polish Academy of Sciences, Rektorska 4, PL-00614 Warsaw, Poland\\ \vspace{2mm} Faculty of Physics, University of Warsaw, Pasteura 5, PL-02093 Warsaw, Poland}

\date{}

\begin{document}

\maketitle

\begin{center}
\textbf{Abstract}
\end{center}

A numerical agent-based spin model of financial markets, based on the Potts model from statistical mechanics, with a novel interpretation of the spin variable (as regards financial-market models) is presented. In this model, a value of the spin variable is only the agent's opinion concerning current market situation, which he communicates to his nearest neighbors. Instead, the agent's action (i.e., buying, selling, or staying inactive) is connected with a change of the spin variable. Hence, the agents can be considered as \emph{cunning} in this model. That is, these agents encourage their neighbors to buy stocks if the agents have an opportunity to sell them, and the agents encourage their neighbors to sell stocks if the agents have a reversed opportunity. Predictions of the model are in good agreement with empirical data from various real-life financial markets. The model reproduces the shape of the usual and absolute-value autocorrelation function of returns as well as the distribution of times between superthreshold losses. \\


\section{Introduction}
\label{intro}

Agent-based modeling plays a significant role in description of financial markets \cite{tesfatsion, arthur, lebaron}. Many important research works on this issue were made at the turn of the century \cite{LM, farmer, Iori, bouchaud, GBM, izumi_ueda, jefferies, lebaron_emp, raberto, tay, kozlowska}. A canonical example is the model based on the concept of mutual exchange and interaction between different groups of investors and the process of price adjustments with a demand-supply imbalance \cite{LM}. Similarly, two types of market participant are assumed in the model \cite{Kaizoji} which is useful to describe the price changes of an asset in short intervals, e.g., during a single day. 

Among later works, an approach based on an assumed market strategy undertaken by the agents is one of the most popular \cite{farmer_joshi, hommes, GB, mizuno, feng, thurner}. Some other are, e.g., approach based on interaction between agents \cite{alfarano} or order book mechanics \cite{preis, chiarella, tedeschi}. 

\subsection{Financial markets and statistical physics}

According to \cite{stauffer2008social} a close connection between the two-dimensional Ising model (or, equivalently, Potts model \cite{wu1982potts}) known from statistical physics, and the socio-economic models exists. Therefore, the works based on different variants of the Potts model are particularly fruitful \cite{LM, Iori, Kaizoji, Sornette, ZS, DGK, takaishi2005simulations, Axel_ABM, epstein2006generative, bourgine2013cognitive, ecnph_review}. In \cite{Iori} an Ising-based model on a square lattice of interacting agents is given, where the connections between agents (spins) are the social connections between the investors, and a threshold mechanism is included in the microdynamics of the model. This model is based on the Potts model, with three possible states of the spin variable, ''buy'', ''sell'' and ''inactive''. Another example of the three-state model is provided in \cite{takaishi2005simulations}. 

Finally, an interesting example of the financial application of the Ising model is presented in \cite{jain2006persistence}. The time evolution of the values of the shares quoted on the London Financial Times Stock Exchange 100 index (FTSE 100) was mapped onto Ising spins. Then, the Ising model simulation provided the evidence for power law decay of the proportion of shares that remain either above or below their 'starting' values. 

Although an overall description of financial markets does not exist and their properties are not purely understood yet, those works give hope for a more comprehensive description of financial markets.

\section{Motivation and basic definitions}
\label{CA_intro}

The model presented in this paper developed from the threshold model of financial markets by Sieczka and Hołyst \cite{SH}. The present model essentially repairs it to make it much more suited for the description of real markets. The reasons why the Sieczka--Hołyst (SH) model is so inspiring are as follows (cf.\ \cite{DGK, MDphdthesis}):
\begin{itemize}
\item[(i)] It contains a threshold mechanism, which is well-established from psychological point of view.
\item[(ii)] It considers the interaction between agents together with the uncertainty in their activity caused by informational noise.
\item[(iii)] It assumes the strength of this interaction, which strongly depends on the macroscopic state of the system.
\item[(iv)] The SH model is sensitive to concrete types of emotion which agents subordinate during the stock market evolution \cite{CCH, KD}.
\end{itemize}

Nevertheless, the SH model contains a fundamental inconsistency, namely, buying (or selling) the stocks by the agents does not lead to change of price. There is even the extreme possibility that all the agents can buy (or sell) the stocks at a given time step, but the corresponding price does not change at all. The purpose of this work is to fix this problem, because, apart of the inconsistency mentioned above, the SH model is very promising. This was an incentive to make its essential modification.

Thus, the motivations to create a new model of financial markets were as follows:
\begin{enumerate}
\item A need to possess a model, as simple as possible, that will reproduce all basic empirical stylized facts, i.e., the volatility clustering of the returns, the shape of the statistics of returns and absolute returns, the behavior of the autocorrelation function of returns and absolute returns.
\item A need to describe numerically a recently discovered stylized fact: universality of the interevent-time distribution on financial markets \cite{BB1, BB2, LTB, LB} (a thorough analytical description of that universality was provided in \cite{DGKJS}).
\item A question if (and how) different types of emotions, that is, different characteristics of the noise in the model change the behavior of the simulated market. 
\item An essential repair of the very promising Sieczka-Hołyst threshold model of financial markets.
\end{enumerate}

Actually, the SH model is based on the standard approach, where the spin variable is interpreted directly as an investment decision, e.g., the purchase of a certain number of stocks. That purchase repeats at the successive time step, even though this variable remains unchanged. The changes of this variable, in turn, could be interpreted as another decision being executed at the successive time steps, e.g., selling stocks. The spin variable in the SH model (and other similar models) means also an opinion about the market communicated to other investors. Due to this communication, the investors mimic the behaviour of their colleagues, i.e., their neighbours in the social net. To sum up, in the standard approach a communicated opinion is identified solely with the (temporary) investment decision, not the (durable) market state of an investor (cf.\ \cite{ZS}).

Conversely, in the model presented in this work, the price can change if and only if the value of spin is \emph{changed}. That is, the investment decision in the model is defined as the \emph{change} of the spin variable of a given agent. We prove below (Sec.\ \ref{CA_desc}) that this protects the model from the inconsistency mentioned above.

This approach appears to be more intuitive and reasonable, as the actual state of a spin is now identified with the actual market state of the corresponding agent, while the action (investment decision) is connected with the change of the state, as expected. Actually, such a property characterizes some epidemic models where the contagion is a change in the state of an agent, from susceptible to infected \cite{epidemic1, epidemic2}, or the models of cultural patterns where the pattern adoption also requires a change of state \cite{axelrod, raducha}. Nevertheless, to the best of our knowledge, this approach is novel in the frame of Potts-based econophysical modeling (cf.\ refs.\ in Sec.\ \ref{intro} above and Sec.\ 5 in \cite{ecnph_review} and refs. therein). 

What is more, the irrational component in the model is provided by the (discrete) Weierstrass--Mandelbrot noise \cite{RK} as opposed to the (continuous) Gaussian one that is frequently used for this purpose \cite{ZS, SH}. This allowed to model some stronger emotions of agents occurring together with gentler ones. The details of the approach sketched in this section and the assumptions of the model are given in Sec.\ \ref{CA_desc}.

The question arises of whether the approach sketched above allows to reproduce the main stylized facts from financial markets, in particular, recently discovered the interevent-time statistics, more carefully examined in \cite{DGKJS}. In Sec.\ \ref{CA_resul} it is shown that the numerical simulations of the model give results consistent with a wide range of empirical financial-market data. Finally, in Sec.\ \ref{CA_sum}, a summary and suggestions of future work are given.

The work in this article was adapted from the doctoral thesis of the author \cite{MDphdthesis}, with appropriate supplemental material added. 

\section{Model description} 
\label{CA_desc}

\subsection{Basic assumptions}

In the present model, we consider $N$ agents (or traders) on a square lattice $n \times n$, $N = n^2$. Each agent is represented by a three-state spin variable $s_i = 0, \pm 1$, $i = 1, \ldots, N$. The social network in the form of a square lattice with a fixed size and only one type of traders on the market are assumed to increase simplicity of the model and check if a more complicated structure is really needed to reproduce the desired market behavior. 

Most directly, $s_i$ is interpreted as the current market state of the $i$th agent, $+1$ when it has bought some stocks (that is, it has the long position open), $-1$ when it has sold some stocks (that is, it has the short position open), and $0$ if it is neutral. We can also consider this value of $s_i$ as advice that the $i$th agent gives to its nearest neighbors at a given time step, when they ask it whether to buy stocks or sell them. Value $s_i = +1$ is the advice to buy, $s_i=-1$ is the advice to sell, while $s_i = 0$ is simply a lack of advice or the advice to wait or stay inactive. Taking into consideration the above comments, we may conclude that the agents are imitating themselves, in an affirmative sense, that is, by giving mutual examples to themselves.

On the other hand, when, for instance, $i$th agent advises the others to buy stocks ($s_i=+1$), it cannot buy them by itself, as the value of its spin variable cannot increase, but can only decrease or remain unchanged. And vice versa for $s_i=-1$, that is, the agent who advises others to sell is only able to buy them now or at least remain inactive. For this reason, the model was called \emph{the model of cunning agents} \cite{DGKA}. The effect mentioned above ensures a negative coupling between subsequent values of a given spin variable. However, for $s_i=0$ the agent can both buy or sell stocks, since it occupies a neutral position. 

In each time period $t$, a spin is drawn a predetermined number of times. The drawn spin, $s_i(t), i=1,2,\ldots,N$, is updated according to the social impact rule
\begin{equation}
s_i(t)=\mathrm{sgn}_{\lambda |M(t-1)|}\left[\sum_{j=1}^{N}J_{ij}s_{j}(t)+\epsilon_i(t)\right],
\label{s_i(t)}
\end{equation}
where
\begin{equation}
\textrm{sgn}_{q}(x)=\left\{ \begin{array}{lll}
+1 & \textrm{if} & x \geqslant q,\\
0 & \textrm{if} & -q \leqslant x < q,\\
-1 & \textrm{if} & x < -q.
\end{array} \right.
\label{sign}
\end{equation}
and the constant $\lambda$ is positive. The exchange coefficient $J_{ij} = J > 0$ if $j$ is one of the four nearest neighbors of $i$ and $J_{ij} = 0$ otherwise, to ensure the (nonnegative) nearest-neighbor interaction. The noise term $\epsilon_i(t)$ is added to include some individual opinions or emotions of agents that accompany their decisions.

The magnetization, $M(t)$, is given conventionally by the following average,
\begin{equation}
M(t)=\frac{1}{N}\sum_{i=1}^{N} s_{i}(t),
\label{MN}
\end{equation}
and indicates the overall market state.

The threshold value $\lambda |M(t)|$ in Eq.\ (\ref{s_i(t)}) is a crucial one with regard to the model mechanics. If the impact of neighbors plus the opinion/emotion of the $i$th agent is lower than this value, the agent switches to neutral $s_i = 0$. A large value of magnetization, corresponding to the large dominance of one of the two -- short or long -- positions occupied by the agents, at fixed $\lambda$ yields a large threshold value and a high probability of switching to neutral position. In other words, the suspicion of  the agents is high in this case. This is understandable, since such a strong imbalance on the market is rather considered as dubious and unsafe.

The spin value $s_i$ is updating according to Eq.\ (\ref{s_i(t)}) immediately, i.e., its neighbors ''see'' its new value just after it was drawn, and also the magnetization value is updated immediately. Moreover, a \emph{round} is defined as $N$ spin drawings, enabling each trader on average one chance to change its state and forming a single time step. 

Notably, Eq.\ (\ref{s_i(t)}) has a completely different interpretation from all the counterparts used earlier (cf.\ \cite{Iori, Kaizoji}). Namely, this formula concerns the \emph{state} of an agent, not its \emph{activity}. Regarding the activity of an agent, it is defined as a \emph{change} of its spin, that is, $d_i(t) = s_i(t) - s_i(t-1)$ is an agent demand for $d_i > 0$ or supply (a negative demand) for $d_i < 0$. Therefore the agent's activity consists of two subsequent different values of the spin variable $s_i$. The agent declares a demand when $s_i$ increases during a given round, or alternatively offers a supply when $s_i$ decreases. This change, $d_i$, can take magnitude 1 (e.g., for a jump from 0 to $-1$) or 2 (for a jump from $-1$ to $+1$ or from $+1$ to $-1$), which is its largest possible value and simultaneously the largest possible single-agent supply or demand for the stocks in a single round. As an assumption, the agent's supply or demand is realized at once, in the same round, and the given agent buys or sells some stocks.\footnote{In a real-life situation, a purchase of a stock is always associated with the sale of another stock of the same asset by another investor. In the model, that view is simplified by neglecting this feature, i.e., an unlimited market liquidity is assumed or only a part of the whole market is considered.}

Evidently, after the purchase or the sale, the agent becomes an advocate of the decision it made and may encourage its neighbors to do the same. Actually, it is also in its best interest, as the one who purchased wants the price to increase, and this is realized when more agents buy, and vice versa in the case of a sale. In this way the mechanism of mutual imitation works. The cycle of the agents' activity and communication is repeated.

To sum up, the model corresponds to the real-world, where investors buy or sell stocks imitating the past behavior of their colleagues, and the more balanced the market situation is, the more willingly they do it. For a case of simplicity, possible demand or supply of a single agent is reduced to one of two elemental magnitudes, equal 1 or 2. 

This reinterpretation of the spin variable prevents the model market from the inconsistency mentioned in Sec.\ \ref{CA_intro}, where the price does not change, although all the agents buy (or sell) stocks. This situation, when all spins have the value $+1$ or all spins have the value $-1$ means now that all investors have, respectively, long or short position open, so the price is not going to change in this case (cf.\ Sec.\ \ref{explanations}), as expected. Additionally, this approach provides a desired correlation between the return volatility and the trading volume in the model. Another essential consequence of this approach is that the threshold slows down both the fast increase and decrease of the price, providing a specific price damping, which cannot itself induce a change in the trend, nor even any oscillations.

\subsection{The noise in the model}
\label{noise}

In the simulations of the model a noise from the Weierstrass--Mandelbrot (WM) probability distribution is used (cf.\ \cite{RK} and refs. therein):
\begin{eqnarray}
p(x) & = & \left(1 - \frac{1}{K} \right) \sum_{j=0}^{\infty} \frac{1}{K^j}\cdot\frac{1}{2}\delta(|x| - b_0 b^j), \nonumber \\ && K, b > 1, \ b_0 > 0.
\label{WM}
\end{eqnarray}
The reasons for using this distribution are as follows:
\begin{itemize}
\item[(i)] Emotions on the market are sometimes very strong; therefore the power-law tails of WM distribution may be more useful to describe these emotions than a simple Gaussian one.
\item[(ii)] WM distribution enables the control of its variance and exponent. The variance may take varied finite values and also be infinite; the distribution may take both L\'evy and non-L\'evy forms asymptotically.
\end{itemize}
Apparently, some other distributions may also possess the qualities (i) and (ii), but WM was picked herein as a reference case (see Sec.\ \ref{CA_sum} for more explanations). WM distribution is a discrete one, having spikes at $x = \pm b_0 b^j, \; j = 0, 1, 2, \ldots$ Its variance is given by
\begin{equation}
\sigma^2 = b_0^2 \frac{1 - 1/K}{1 - b^2/K} \ \textrm{for} \ b^2/K < 1;
\label{sigma2}
\end{equation}
otherwise, it is infinite.

We can easily prove that for $|x| \gg 1 / \ln{b}$ Eq.\ (\ref{WM}) takes the approximated form
\begin{equation}
p(x) = \frac{1-1/K}{\ln{K}} \frac{b_0^{\beta}}{|x|^{1+\beta}},
\label{appWM}
\end{equation}
where $\beta = \frac{\ln{K}}{\ln{b}}$, and the variance of the above given approximate PDF is finite only for $\beta > 2$; otherwise, it is infinite.

Due to the form of the WM distribution, it always provides some, at least small, emotions for the agents behavior, cf.\ Eq.\ (\ref{WM}). What is more, although Eq.\ (\ref{appWM}) represents a power-law distribution, only the part of it nonexceeding the maximal range $[-(4J + \lambda); 4J + \lambda]$ is essential for the dynamics of the system. Inside this range, spin value $s_i$ can vary between $-1, 0$, and $+1$. Above its right border, $s_i$ can only take value $+1$, while below its left border it can only take value $-1$. The reason for such behavior is the threshold character of the social impact function (\ref{s_i(t)}).

\subsection{Some additional formulas and assumptions}
\label{explanations}

After \cite{RJB}, usually considered in ABMs is the logarithmic return, $r$, proportional to the properly defined excess demand $D$ (cf.\ \cite{LM, Gont}). This feature can be written as
\begin{equation}
r_{\tau}(t) = \ln{P(t)} - \ln{P(t-\tau)} = \frac{1}{\Lambda} D_{\tau}(t),
\label{ror}
\end{equation}
where $P$ is price,
\begin{equation}
D_{\tau}(t) = \sum_{i=1}^{N}[s_{i}(t) - s_{i}(t-\tau)] = N[M(t) - M(t - \tau)],
\label{rortau}
\end{equation}
and $\tau$ is delayed time in units of rounds (for $\tau = 1$, $r_1 \equiv r$), while the coefficient $\Lambda$ can be interpreted as the market depth. Actually, taking a particular value of time lag $\tau$ is arbitrary, since the system is updating itself after each spin change (see Sec.\ \ref{CA_desc}). It is worth noting that the $D_{\tau}(t) \neq 0$ only if the state of at least one agent changes. Naturally, the changes of states may cancel each other, leaving the excess demand unchanged. Thus, the price $P$ changes if and only if the magnetization $M$ changes, i.e., when there is an imbalance between the overall demand and supply on the market, as in a real-life situation.

To cope with an occasional trapping of the system into the fully ferromagnetic state, when all agents have the same value of their spin variables, either $+1$ or $-1$,\footnote{This can be interpreted as a fully illiquid market; cf.\ \cite{LM}.} an exogenous factor is introduced to act as a market maker. It abruptly restores the system to a paramagnetic state with randomly oriented spins. After that, the computations are continued until the next such restart.\footnote{The convergence to a fully ordered state is not a question of preponderance of one of the fractions ($-1$, 0, or $+1$). When there is an even drawing for each of these values, there are still many restarts. See \cite{trap_avoid} for some other attempt to avoid such ordering behavior in the Ising model.}

The model described above is as simple as possible, therefore it is only an approximation of a real situation, where each agent can buy and sell only alternately. We can make a simplifying assumption that each purchase or sale made by an agent is in fact the sum of all its purchases or sales before he changes his/her decision into opposite (i.e., sale or purchase); however, the advice he/she is giving to his/her neighbors remains the same in this case, which is a correct behavior. Despite that, the restarts of the system, described above, partly allow agents to buy after the previous purchase and to sell after the previous sale and, as a result, they allow the greater price increases.

\section{Results and empirical data comparison}
\label{CA_resul}

\subsection{Basic results}
\label{CA_basic}

\begin{figure}
\begin{center}
\includegraphics[width=\textwidth]{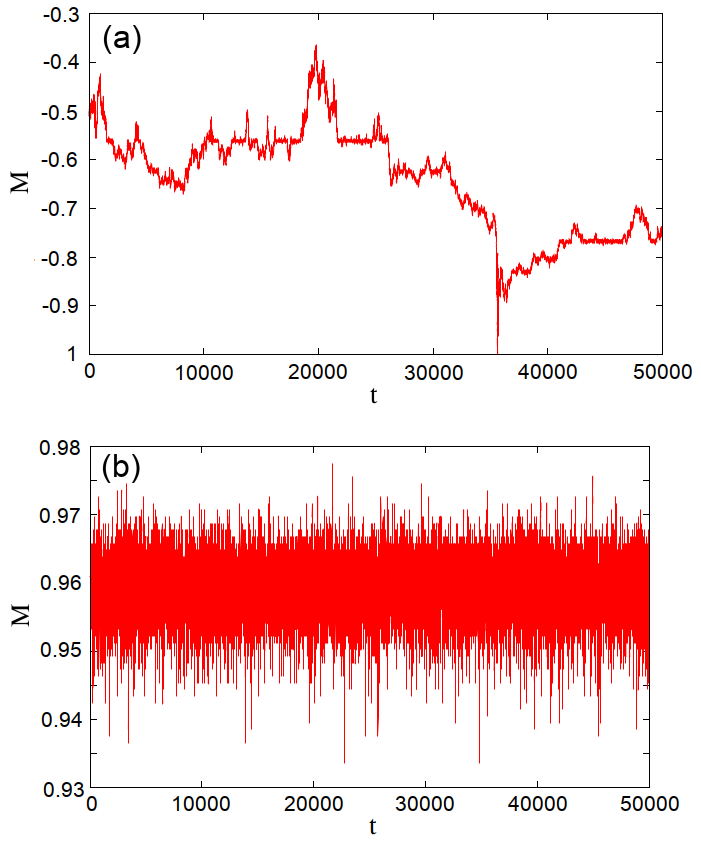}
\caption{Phase transition of the magnetization evolution characteristics for parameters $J=1, K=5, b=2, b_0=0.2$, and (a) $\lambda = 2.2$, (b) $\lambda = 2.3$.}
\label{transit}
\end{center}
\end{figure}

After \cite{SH} the lattice size $N = 32 \times 32 = 1024$ agents is set and at the beginning of the simulation a random configuration of the spin values is drawn. For an increasing value of the threshold parameter $\lambda$, a phase transition between walk- and noise-like behavior of the magnetization evolution occurs (cf.\ Fig.~\ref{transit}).

\begin{figure}
\begin{center}
\includegraphics[width=\textwidth]{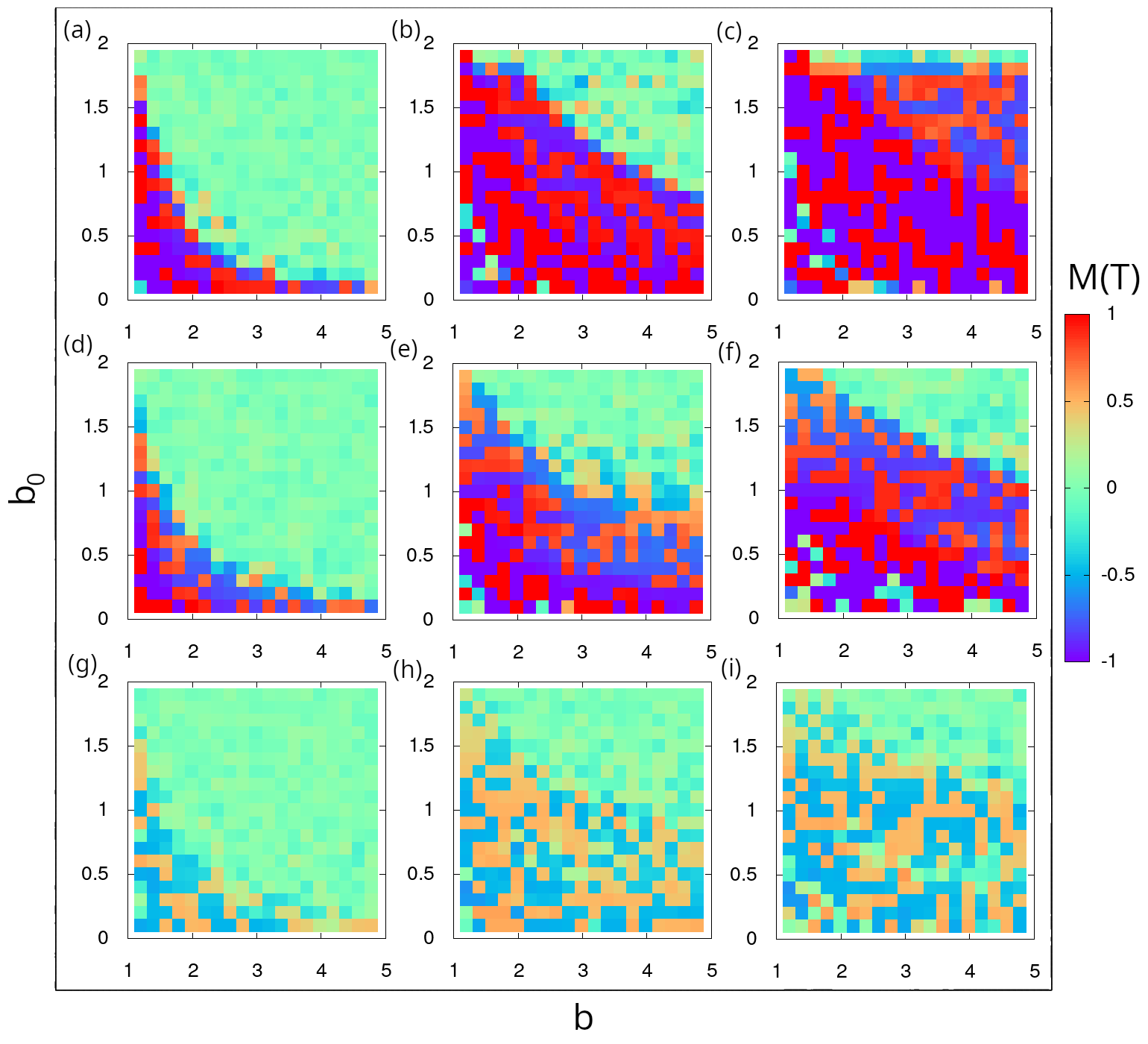}
\caption{Representative phase diagrams of $M(T)$ (color scale) for $T = 100~000$, different values of $b$ (horizontal axis) and $b_0$ (vertical axis), and for (a) $K = 2, \lambda = 0.3$, (b) $K = 5, \lambda = 0.3$, (c) $K = 8, \lambda = 0.3$, (d) $K = 2, \lambda = 2.1$, (e) $K = 5, \lambda = 2.1$, (f) $K = 8, \lambda = 2.1$, (g) $K = 2, \lambda = 3.9$, (h) $K = 5, \lambda = 3.9$, (i) $K = 8, \lambda = 3.9$.}
\label{pd}
\end{center}
\end{figure}

The question appears, what the overall behavior of the system is, for a whole spectrum of the parameters $J, \lambda, K, b$, and $b_0$. For this purpose, a large simulation of the model for $\lambda = 0.1, 0.3, \ldots, 3.9$, $K = 1.5, 2, \ldots, 9.5$, $b = 1.2, 1.4, \ldots, 4.8$, and $b_0 = 0.1, 0.2, \ldots, 1.9$ was performed (as parameter $J$ is redundant and can be absorbed defining $\lambda' = \lambda/J, \epsilon'_i = \epsilon_i/J, J' = 1$). A unit simulation for each set of the parameters within their ranges and steps given above, lasted $T = 100~000$ rounds, the (arbitrarily taken) time sufficient for the system to jump to its long-term state, in terms of the absolute value of the system's magnetization $M(T)$ (see below). The representative results of these simulations, in the form of two-dimensional phase diagrams --- ''slices'' of the four-dimensional phase space --- are shown on Fig.\ \ref{pd}. All other results looks similar, and can be extrapolated from those presented on Fig.\ \ref{pd}. 

Apparently, all plots are symmetrical in terms of positive or negative value of $M(T)$, as expected (cf.\ Sec.\ \ref{CA_sum}). Qualitatively, there are three possible long-term states of a final magnetization value: (i) the value oscillating around zero (paramagnetic state), (ii) the absolute value somewhere between zero and one (ferromagnetic state), and (iii) the absolute value (almost) equal to one (fully ferromagnetic state). The greater the value of $b$ or $b_0$ is, the more probable is the paramagnetic state, which is understandable, because both parameters are proportional to the noise strength in the model. On the contrary, increasing $K$ value (from the left to the right of Fig.\ \ref{pd}) makes the ferromagnetic state more probable. Indeed, the increase of this parameter makes the high values of the noise less probable [cf.\ Eq.\ (\ref{WM})] and therefore increase the probability of the ordered state of the system. Finally, parameter $\lambda$ (increasing from the top to the bottom of Fig.\ \ref{pd}) is responsible for switching between fully (low $\lambda$) and non-fully ferromagnetic state (high $\lambda$), as the threshold value in Eq.\ (\ref{s_i(t)}) is proportional to $\lambda$.

\begin{figure}[t]
\begin{center}
\includegraphics[width=\textwidth]{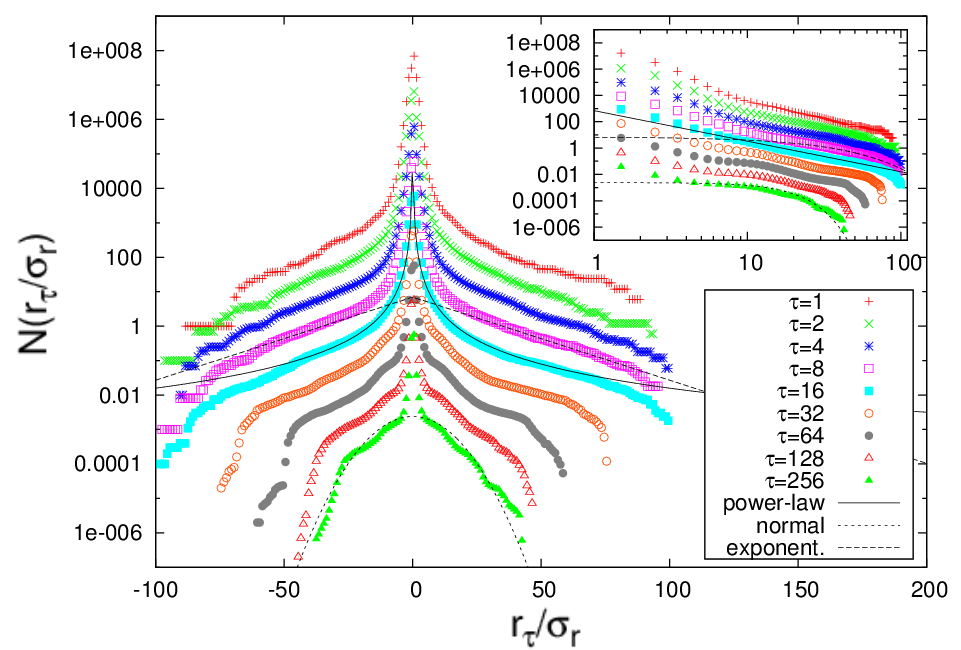}
\caption{Numerical statistics of returns $r_{\tau}(t)$ for $J=1, \lambda=1, K=5, b=2, b_0=0.2$, and different $\tau$ (see the legend). For viewing purposes, the returns are rescaled by their standard deviations and the histograms are shifted upward with successive powers of 10. Power-law (solid line), Gaussian (denoted as normal; dotted line) and exponential (dashed line) functions fitted to selected histograms are shown.}
\label{stat}
\end{center}
\end{figure}

Return statistics from the model simulation reveal power-law tails for varied values of time lag $\tau$, as in Fig.\ \ref{stat}.
Because the noise, $\epsilon_i$, is cut off by values $-(4J+\lambda)$ on the left and $4J+\lambda$ on the right [cf.\ Eqs.\ (\ref{s_i(t)}) and (\ref{sign})], for parameters as in Fig.\ \ref{stat} only components with $j = 0,1,2,3,4$ are significant values in the sum in Eq.\ (\ref{WM}). For most values of $\tau$, the distributions reveal explicit power-law tails with the exponent close to 3, even despite the aforementioned cut-off of the noise and the fact that the range $[-(4J+\lambda), 4J+\lambda]$ is narrower than its equivalent in the SH model. Only for large values of $\tau$ does the power-law character of the statistics break down, nevertheless, fat tails still occur, as opposed to the SH model, where the authors obtained convergence to Gaussian tails for the largest values of $\tau$. This is a strong argument that the power-law behavior is not driven by a type of noise assumed in the model. 

\begin{figure}[h]
\begin{center}
\includegraphics[width=\textwidth]{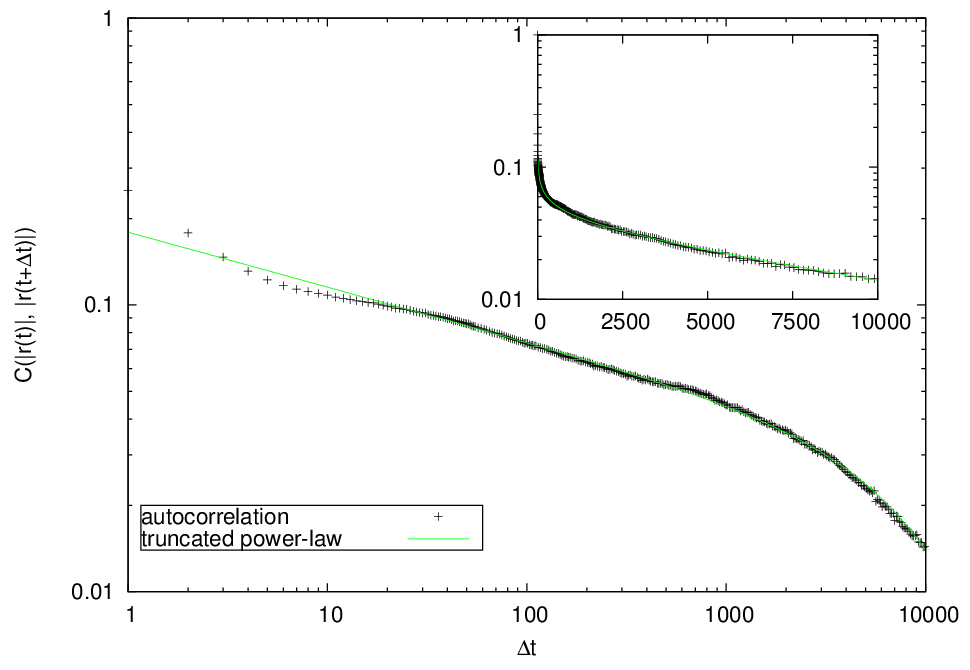}
\caption{The autocorrelation function of absolute returns for $J=1, \lambda=1, K=5, b=2, b_0=0.21$ on log-log and (inset) lin-log scale (black pluses). The green solid line is the fit of an exponentially truncated power-law function.}
\label{auto}
\end{center}
\end{figure}

The autocorrelation function\footnote{Due to the fact that time in the model of cunning agents is discrete (measured in units of rounds), the autocorrelation function is defined as $C(x(t),x(t+\Delta t)) = \frac{\left\langle (x(t) - \mu)(x(t + \Delta t) - \mu) \right\rangle}{\langle (x(t) - \mu)^2 \rangle};\ \mu = \langle x(t) \rangle$. Thus, the length of time intervals between subsequent returns is not taken into consideration, herein.} of absolute returns in the model are slowly decaying, and they frequently resemble an exponentially truncated power-law function (cf.\ Fig.\ \ref{auto}), which is consistent with empirical findings \cite{arneodo, raberto_emp, lillo_farmer, cont}.

\subsection{Empirical-data comparison}
\label{CA_empir}

\begin{figure}
\begin{center}
\includegraphics[width=\textwidth]{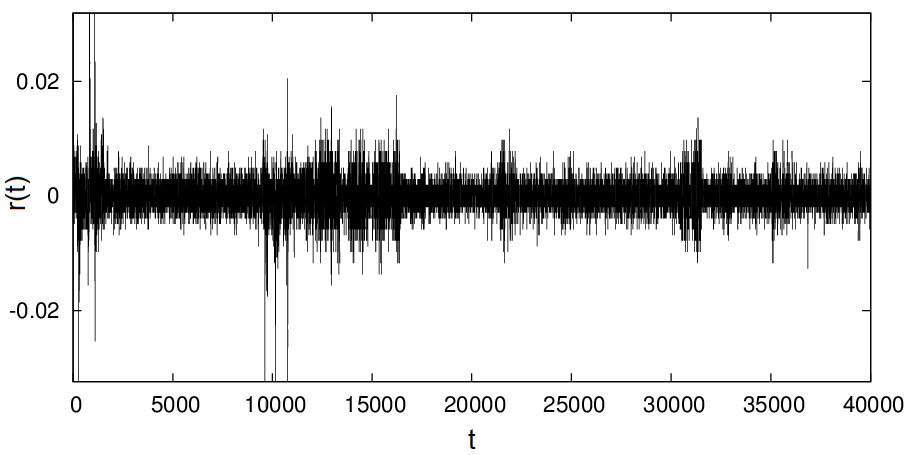}
\caption{Returns in time for an example run of the cunning-agents model simulation for the parameters $J=1, \lambda=1, K=5, b=2, b_0=0.2$. Volatility clustering (changes of the signal's standard deviation in time) is visible.}
\label{returns}
\end{center}
\end{figure}

\begin{figure}
\begin{center}
\includegraphics[height=0.75\textheight]{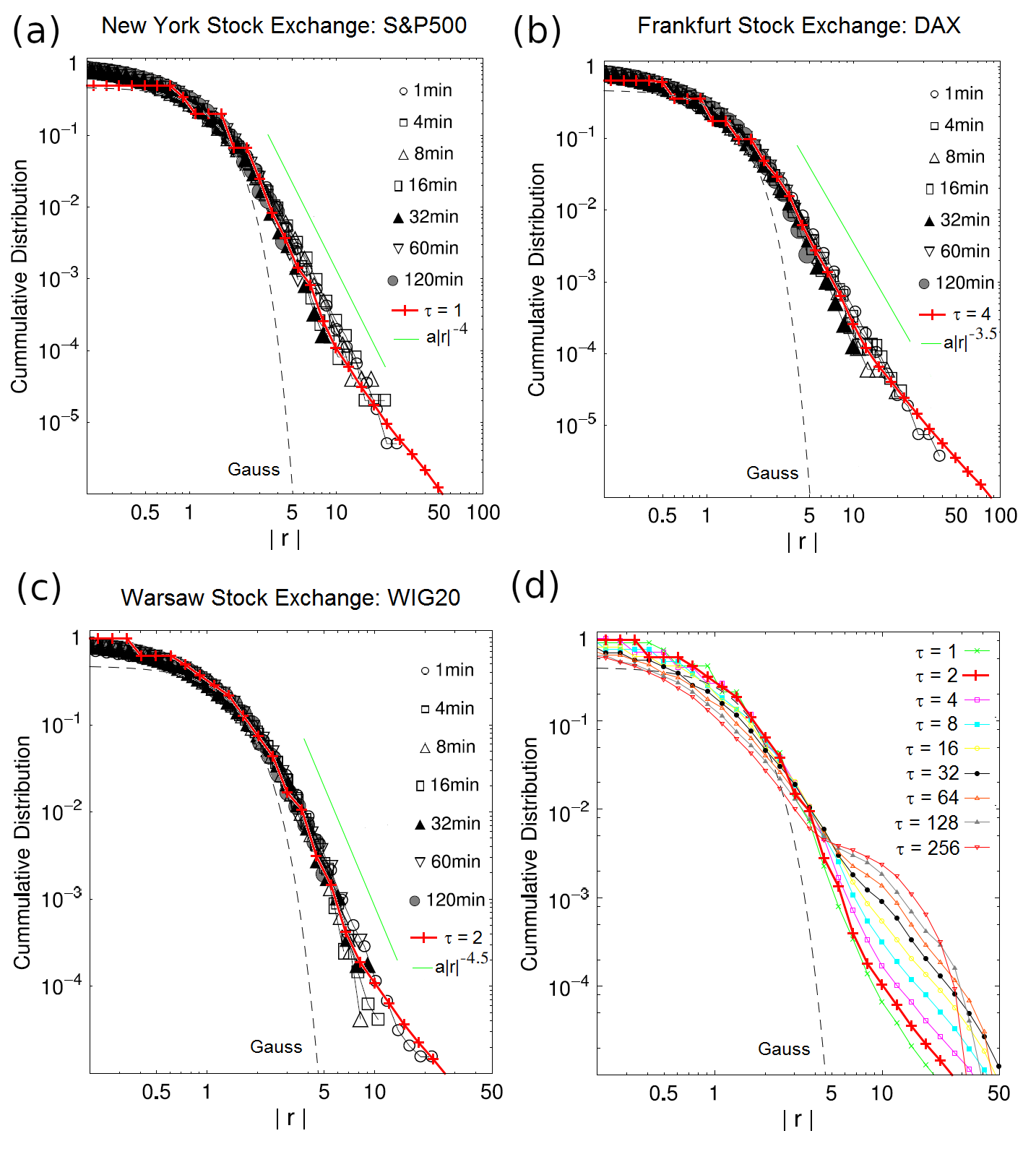}
\caption{(a)-(c) A comparison between cumulative distributions of absolute returns for the model and real-market data (matching by the trial and error method with a fit accuracy of about $10\%$). The theoretical results are shown by markers connected by segments (in red), while the empirical results are denoted by markers connected by dashed curves (in black). (a) For the American S\&P500 index, $J = 1$, $\lambda = 2.2$, $K = 6.5$, $b = 2$, $b_0 = 0.2$, $\tau = 1$. (b) For the German DAX, $J = 1$, $\lambda = 2.1$, $K = 6.5$, $b = 2$, $b_0 = 0.2$, $\tau = 4$. (c) For the Polish WIG20, $J = 1$, $\lambda = 2.2$, $K = 5$, $b = 2$, $b_0 = 0.25$, $\tau = 2$. The green straight lines in plots (a)-(c) denote a power law driven by exponents equal to 4.0, 3.5, and 4.5, respectively. (d) For instance, the model statistics for different values of $\tau$ (see the legend for details) but for other parameters the same as in (c). Dashed single curves show Gaussian predictions. Empirical data reprinted from \cite{Drozdz} \textcopyright 2007, with permission from Elsevier.}
\label{statemp}
\end{center}
\end{figure}

\begin{figure}[t]
\begin{center}
\includegraphics[width=0.9\textwidth]{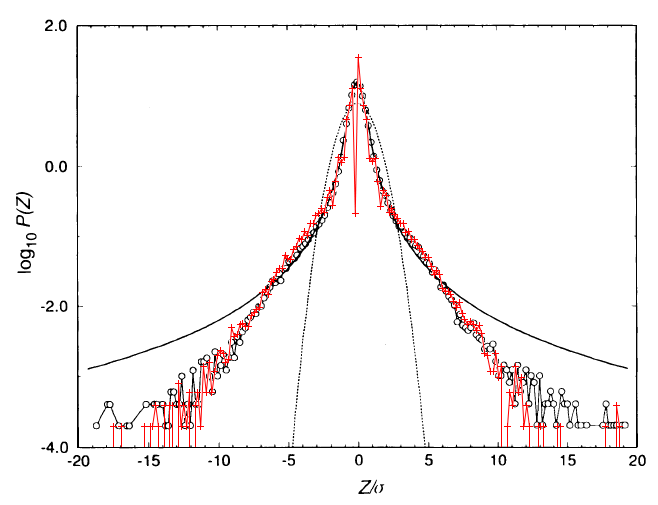}
\caption{The distribution of one-minute returns for the S\&P500 index from January 1984 to December 1989 (black empty circles and solid segments) and the model returns $r_1$ for 500~000 time steps, $J=1, \lambda=1, K=5, b=2, b_0=0.2$ (red pluses and solid segments). Additionally, the solid line shows symmetrical L{\'e}vy stable distribution, while the dotted line shows Gaussian distribution. It can be seen that the empirical distribution is reproduced by the model in the entire range of returns (matching was done by the trial and error method, with the fit accuracy of about $10\%$). Adapted by permission from RightsLink: Springer Nature NATURE, \cite{MS} \textcopyright 1995.}
\label{MS_fig}
\end{center}
\end{figure}

As was indicated in Sec.\ \ref{CA_intro}, the model is capable of reproducing a wide range of empirical results. Firstly, return variograms with volatility clustering resemble the real-market ones (Fig.\ \ref{returns}).
Secondly, cumulative absolute-return distributions for stock markets of essentially different capitalizations, with varying slope of tails are reproduced quite well by the model of cunning agents (by varying few driven parameters; Fig.\ \ref{statemp}).
As presented in Fig.\ \ref{statemp}(d), the model distributions show some nontrivial structures for larger values of time lag $\tau$. Fortunately, they agree quite well with the empirical data within their attainable range. A distribution comparison for a full range of the return values (i.e., positive and negative) is shown in Fig.\ \ref{MS_fig}.
It is evident that the model of cunning agents reproduces both the absolute and the usual market-return statistics.

\begin{figure}[t]
\begin{center}
\includegraphics[width=0.9\textwidth]{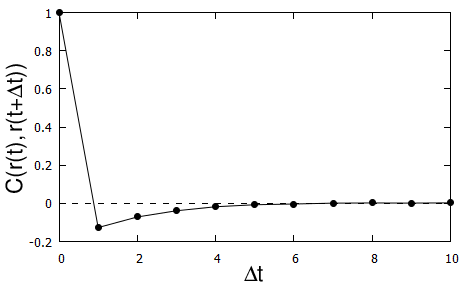}
\caption{The autocorrelation function of returns in the model for the parameters $J=1, \lambda=1, K=5, b=2, b_0=0.2$. The dashed horizontal line stands for ordinate zero.}
\label{auto_usual}
\end{center}
\end{figure}

Also, the behavior of the returns' autocorrelation function is reproduced well within the frame of this approach. In case of raw (not absolute) returns, the empirical autocorrelation function is usually fast decaying with only initial values significant. For short time resolutions (i.e., for so-called tick data) for time lag $\tau = 1$ it is expected to be negative due to the antipersistent character of the financial time series on short time scales, related to the bid-ask bounce effect \cite{arneodo, cont_styl, GK, preis_emp}. Substantially, this behavior characterizes the autocorrelation function from the model shown in Fig.\ \ref{auto_usual}.

\begin{figure}[t]
\begin{center}
\includegraphics[width = 0.9\textwidth]{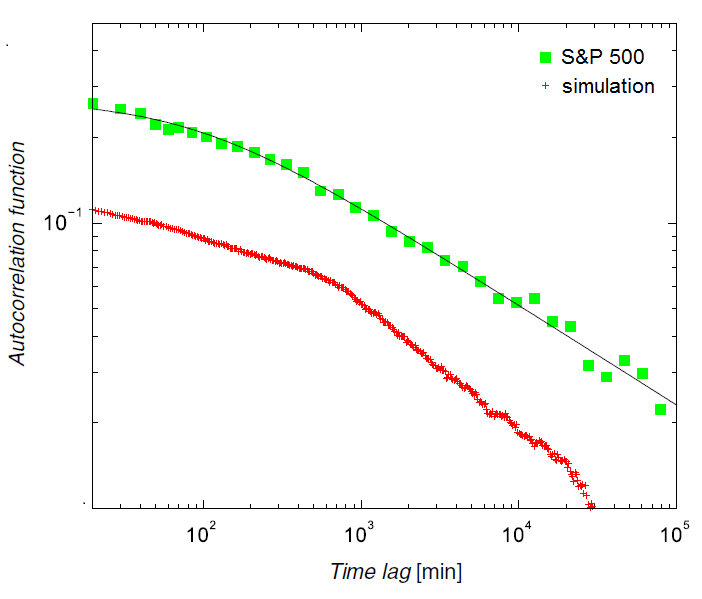}
\caption{The autocorrelation function of absolute returns for S\&P500 (green points) and of $|r_1(t)|$ from the model (red points) for $J = 1$, $\lambda = 2.1$, $K = 6.5$, $b = 2.1$, and $b_0 = 0.2$. Given that the trading day on the NYSE lasts 6.5 h, a single time step, herein (i.e., $\tau = 1$) calibrates as approximately 1 min. Empirical data, including the dashed curve drawn with a rough estimation, reprinted from \cite{Stanley} \textcopyright 2010, with permission from Elsevier.}
\label{autoemp}
\end{center}
\end{figure}

Comparison for the case of the absolute-return autocorrelation function is shown in Fig.\ \ref{autoemp}. Evidently, the shapes of the numerical and the empirical curve are very similar; also the slopes of the tails of both autocorrelation functions agree quite well with each other. Although similar results regarding the usual as well as the absolute-return autocorrelation function were obtained previously by other authors, we have not found the actual reproduction of either of them by an agent-based model, which would comprise the initial significant negative value of the first one as well as the power-law behavior for the second one, as in the model of cunning agents (cf.\ refs.\ in Sec.\ \ref{intro}).

\subsection{Interevent-times description}
\label{CA_iet}

As shown previously in this section, the main stylized facts from financial markets are already reproduced well by the cunning-agents model. Additionally, a comparison of the model predictions with the universal distribution of the interevent times in financial-market data (cf. \cite{DGKA}) is provided herein. The shape of this distribution, i.e., the distribution of times between profits or losses which exceeds some fixed threshold value $Q$, has been proven recently to depend only on this threshold, but not, e.g., on an underlying asset type or the time resolution of data \cite{BB1, BB2, LTB, LB}. 

The analytical superstatistics model, presented in \cite{DGKJS, DJGKS}, provided an analytical formula for interevent-time, $\Delta_Q t$, distribution function:
\begin{equation}
\psi_Q^{\pm}(\Delta_Q t) = \frac{1}{\tau_Q^{\pm}(Q)}\frac{\alpha_Q^{\pm}}{\left(\frac{\Delta_Q t}{\tau_Q^{\pm}(Q)}\right)^{1\pm \alpha_Q^{\pm}}} \Gamma^{\pm}\left(1 \pm \alpha_Q^{\pm},\frac{\Delta_Q t}{\tau_Q^{\pm}(Q)}\right),
\label{iet_form}
\end{equation}
where $\tau_Q^{\pm}(Q)$ and $\alpha_Q^{\pm}$ are parameters, both dependent on the threshold $Q$, while $\Gamma^{\pm}(\cdot,\cdot)$ denotes the lower (for ``$+$'') and upper (for ``$-$'') incomplete gamma functions. Apparently, the above analytic formula is a power-law function corrected by the lower or upper incomplete gamma function for small values of $\Delta_Q t$. For more explanations see \cite{DGKJS}.
\begin{figure}[t]
\begin{center}
\includegraphics[width=0.9\textwidth]{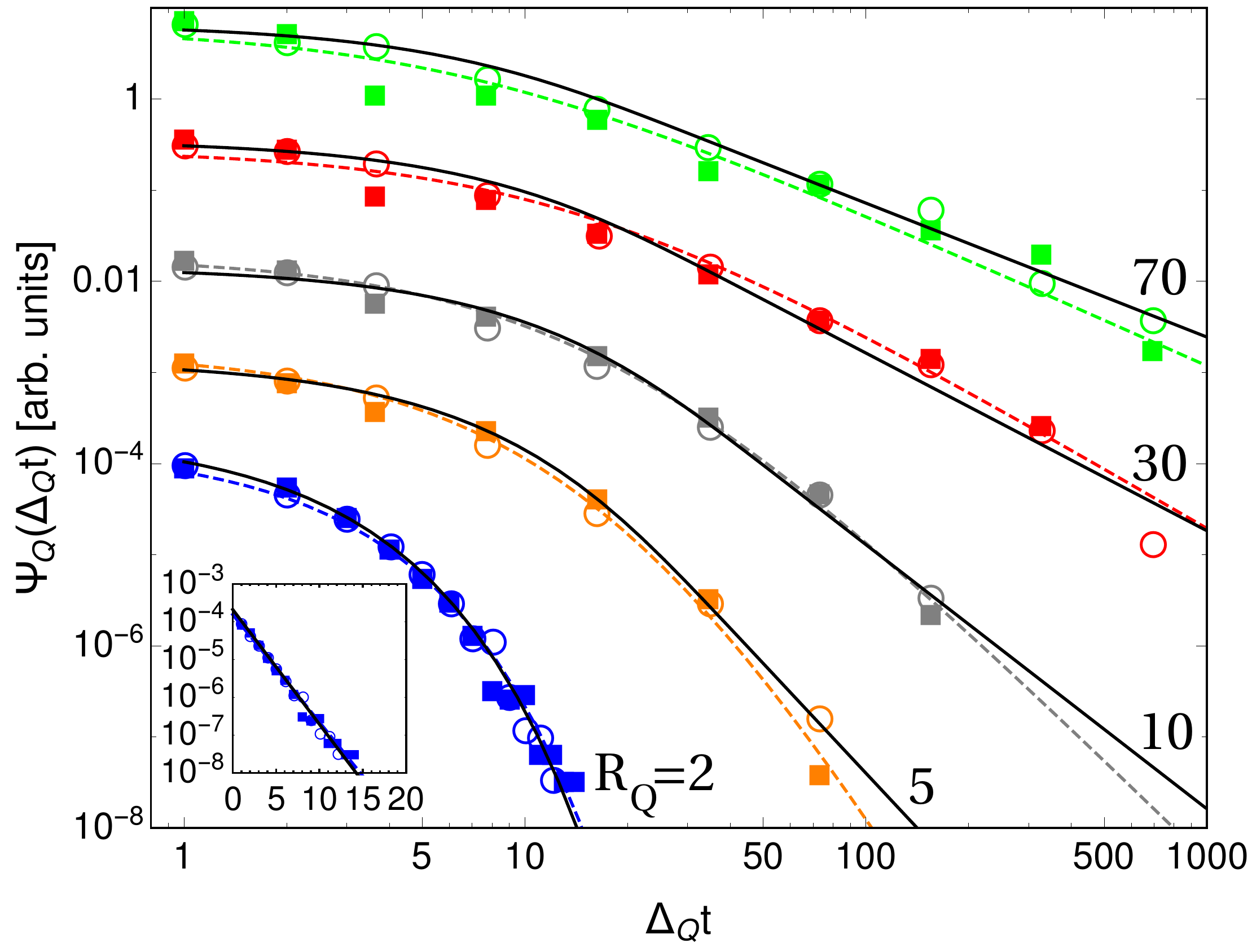}
\caption{Distributions of interevent times, $\psi_Q(\Delta_Q t)$, vs.\ interevent time, $\Delta_Q t$, in log-log scale for example simulation data from the cunning-agents model (full colored squares) for $R_Q=2, 5, 10, 30, 70$ and $\tau = 10^3\, [\textrm{rounds}]$ (estimated equivalent of a single trading day), and for the parameters $J=1, \lambda=2, K=5, b=2, b_0=0.2$ (hence, the Pareto exponent $\beta = \ln K/\ln b > 2$, cf.\ Eqs.\ (\ref{WM}) and (\ref{appWM})). The inset is plot of $\psi_Q(\Delta_Q t)$ vs.\ $\Delta_Q t$ in semilogarithmic scale for $R_Q=2$. Empty colored circles are the empirical data taken from \cite{LTB}. Additionally, the black solid curves are predictions of Eq.\ (\ref{iet_form}) (or Eq.\ (15) from \cite{DGKJS}), while the dashed curves are $q$-exponential fits.}
\label{figure:AO14}
\end{center}
\end{figure}

The problem of excessive losses is significant in economic theory and investment practice, as well as for random processes analysis. Therefore Eq.\ (\ref{iet_form}), which provides the description of the time distance between the subsequent losses of a particular magnitude, has a great importance in the analysis of losses in financial-market time series and is closely related to the economic concept of value at risk. Actually, using Eq.\ (\ref{iet_form}), we may simulate value-at-risk time evolution and predict the time to the next loss exceeding some particular threshold \cite{MDphdthesis, DGKJS}.

It is clear from Fig.\ \ref{figure:AO14} that the numerical predictions of the cunning-agents model agree with the empirical data of interevent times, as well as with the analytical superstatistics model (Eq.\ (\ref{iet_form})). To the best of our knowledge, this comparison, presented initially in \cite{DGKA}, was the first successful attempt at reproducing the universality discovered in \cite{LB} by a microscopic, agent-based model (cf. \cite{Gont}). Furthermore, this is a confirmation of the correctness of the analytical superstatistics model. 

\section{Summary}
\label{CA_sum}

In this paper, a three-state two-dimensional Potts model interpreted as agent-based numerical model of financial markets was presented. Spins in the model, $s_i$, are investors (``spinsons'') imitating their nearest neighbors on a model social network by taking similar market positions, that is, long position for $s_i = +1$, a short one for $s_i = - 1$, and a neutral state for $s_i = 0$. The actual investment decision is in this case the change of the spin variable, $d_i(t) = s_i(t) - s_i(t-1)$, which is intuitively understandable, since the state of a spin indicates, herein, the state of an investor and the action is associated with the change of this state.

In this way, opinion and decision, which hitherto have been generally dealt with together, are separated, and the concepts of opinion and decision modeling are connected within the model of cunning agents. Such an approach is novel in the frame of econophysical modeling; however, it resembles the approach from some sociophysical models, where the spin frequently stands for an agent's opinion, which is rather a state (cf.\ \cite{spin_opinion}). We may conclude that in this approach the decision is more difficult, since it requires change of the state, and thus the model is more ``passive'' than the prevailing market models.

To the best of our knowledge, such interpretation of a spin variable, as a market state, rather than the activity of an investor, is different to the interpretation assumed beforehand in financial-market ABMs (cf. \cite{KD, ecnph_review, Tsallis} and refs.\ in Sec.\ \ref{intro}) and it constitutes the key concept of the model of cunning agents.

The four-dimensional phase diagram for the model reveals the influence of its key parameters on the long-term behavior of prices (cf.\ Sec.\ \ref{CA_basic}). Moreover, the model provides successful empirical comparisons with the usual and absolute-value return autocorrelation, and with other quantitative characteristics of financial-market data, e.g., the  interevent-time statistics (cf.\ \cite{DGKA}), which confirms the legitimacy of this approach. These results are obtained despite using a simple social network in the form of a square lattice and only one type of traders, which appears to be fully sufficient in this case. 

The comparison of the model predictions with the interevent-time empirical data (cf.\ Fig.\ \ref{figure:AO14}) was made without rescaling the relevant statistics. Ipso facto, a mean value of the interevent time is allowed to be infinite, which distinguishes this approach from that used later in \cite{Gont}, and makes it better suited to market reality. 

Furthermore, to take into account the possibility of some intense emotions in the model, the Weierstrass--Mandelbrot noise is used in the agents' activity, instead of the canonical Gaussian one. We verified that, despite the presence of a threshold in the model, the results obtained for Gaussian and Weierstrass--Mandelbrot noises are different. For instance, for the WM distribution we obtain a power-law return and absolute-return statistics for (almost) the whole time-resolution range, while for the normal distribution the corresponding range is shorter. What is more, the discreteness of the WM distribution may also affect the results somehow. Therefore, apart from the WM and the canonical continuous Gaussian distribution, also the discrete Gaussian distribution and the continuous Pareto distribution were examined, and the WM distribution appears to produce the most convincing results from the four distributions used. 

On the other hand, the noise itself cannot be fully responsible for the observed power-law tails, as for the results presented in Sec.\ \ref{CA_resul} the maximal range of the noise-term impact is $[-(4J + \lambda), 4J + \lambda]$, where $\lambda$ is the maximal threshold value, and $J$ is the strength of the nearest neighbors' interaction (cf.\ Sec.\ \ref{noise}). It means a strong cut-off of the noise term, which prevents the strong impact of the tails of the noise distribution on the final results. Therefore the obtained properties of financial market data generated by the model cannot be directly controlled by properties of the noise distribution, but rather by the model mechanics itself.

Extension of this approach to some more realistic topologies of the social network, as well as some other noises assumed, appears to be possible. It would be particularly interesting to incorporate order-book mechanics into the model and to observe its influence on the results. Evidently, the model in the present version is perfectly symmetrical, therefore breaking this symmetry would be desirable. Finally, some other modifications and generalizations of the basic version, adjusting it to the market reality, would be appreciated as well.

Furthermore, the role of abrupt transitions in the model mechanics shall be thoroughly considered. Some reasonable possibility to dispose of them without losing the obtained market similarities would be welcome. Another interesting question is how the presence of the threshold in the model, in particular, the limit for the noise influence on an agent's state established by it, affects the shape of statistics and autocorrelation of returns, e.g., their exponential cut-off at their very end. 

We could also attempt to reconstruct the multifractal spectra of a real-life return time series, as the model has not been analyzed in this matter yet (cf. \cite{GC}). Finally, it would be a challenge to solve the three-state Potts model used in this approach and obtain the result in a closed form. In the long run, the goal for the cunning-agents model would be the Holy Grail of market modeling, namely, the prediction of market behavior, or at least a partial one.

\section*{Acknowledgements}

MD is supported by the grant \emph{AstroCeNT --- Particle Astrophysics Science and Technology Centre International Research Agenda} carried out within the International Research Agendas programme of the Foundation for Polish Science financed by the European Union under the European Regional Development Fund. MD is grateful for inspiring discussion and valuable comments made by Ryszard Kutner and Piotr Gawron. 

\bibliographystyle{unsrt}

\end{document}